\documentclass[letterpaper,journal]{IEEEtran}
\usepackage{graphicx}
\usepackage{amsmath}
\usepackage{amssymb}
\usepackage{epsfig}
\usepackage{epstopdf}
\usepackage{amsthm}
\usepackage{afterpage}
\usepackage{amsmath,amssymb,amsfonts}
\usepackage{verbatim}
\usepackage{psfrag}
\usepackage{color}
\usepackage{cite}
\usepackage{orcidlink}
\usepackage{setspace}
\usepackage{adjustbox}
\usepackage{epsfig}
\usepackage{epstopdf}
\usepackage{footmisc}
\usepackage{fmtcount}
\usepackage{stackrel}
\usepackage{float}
\usepackage{breqn}
\usepackage{hyperref}
\usepackage{enumerate}
\usepackage{graphicx}
\usepackage{algorithm}
\usepackage{algorithmicx}
\usepackage{algpseudocode}
\usepackage{graphicx}
\usepackage[caption=false,font=footnotesize]{subfig}

\usepackage{tabularx,booktabs}

\begin{document}

\title{Resource-Aware Topology Management for ISAC-Enabled TDOA Localization in IoUT Networks}
\author{Ruhul Amin Khalil\orcidlink{0000-0003-4039-9901}, \IEEEmembership{Member, IEEE}
\thanks{This work was supported by the Office of the Associate Provost for Research at the United Arab Emirates University (UAEU), UAE.}
\thanks{R. A. Khalil is with the Engineering Requirement Unit (ERU), College of Engineering, United Arab Emirates University, Al-Ain 15551, UAE (e-mail: ruhulamin@uaeu.ac.ae).}
}

\maketitle

\begin{abstract}
Reliable localization in the Internet of Underwater Things (IoUT) is hindered by limited acoustic bandwidth, long propagation delays, multipath, Doppler shifts, and energy-constrained nodes. This letter proposes an integrated sensing and communication (ISAC)-enabled topology management and multi-stage adaptive estimation (ISAC-TM-MAE) framework for time-difference-of-arrival (TDOA) localization in IoUT networks. In the proposed framework, heterogeneous underwater sensors, seabed anchors, autonomous underwater vehicles (AUVs), and surface buoys exploit acoustic ISAC packets for joint communication and localization. The IoUT network is modeled as a graph, in which informative node pairs are selected subject to acoustic bandwidth, computation, energy, and link-reliability constraints. A D-optimal criterion maximizes the Fisher information matrix (FIM) over the underwater region of interest, while the MAE stage refines the source estimate under acoustic noise and channel uncertainty. Numerical results show that the proposed ISAC-TM-MAE improves localization accuracy, resource efficiency, scalability, and robustness compared with centralized, distributed, and benchmark TDOA localization schemes.
\end{abstract}

\vspace{-2em}

\section{Introduction}
\IEEEPARstart{T}{he} IoUT enables marine applications such as oceanographic data collection, pollution monitoring, offshore inspection, disaster prevention, assisted navigation, and tactical surveillance \cite{khalil2020toward,Jouhari2019UWSNIoUT,khalil2026semantic}. Unlike terrestrial IoT networks, IoUT systems mainly rely on underwater acoustic communication because radio-frequency signals suffer severe attenuation in seawater, while optical links are limited by turbidity, alignment, and short communication ranges. However, acoustic links suffer from low propagation speed, limited bandwidth, long delay, frequency-dependent attenuation, multipath, Doppler spread, and strict energy limitations \cite{jehangir2024isac}. These features make accurate and resource-efficient localization essential for interpreting sensed data, coordinating underwater platforms, tracking targets, and maintaining reliable network operation.

Several underwater localization techniques have been studied, including angle of arrival (AoA), received signal strength (RSS), time of arrival (TOA), TDOA, and frequency difference of arrival (FDOA) \cite{yan2025survey}. Among them, TDOA is attractive for IoUT networks because it estimates the source position from differential arrival-time measurements without requiring tight synchronization between the source and receiving nodes. Recent works have investigated TDOA-based underwater positioning using chirp signals in acoustic channels \cite{Rezzouki2024NetFishingTDOA} and synchronized-transmission TDOA localization for IoUT under depth-dependent sound speed \cite{Chen2025SyncTDOAIoUT}. Nevertheless, underwater TDOA localization remains challenging due to sound-speed variations, node drift, multipath, Doppler effects, and limited acoustic resources. Most existing methods focus on estimator design or propagation modeling, while resource-aware selection of informative TDOA links under communication, computation, and energy constraints remains underexplored.

ISAC enables the sharing of waveforms, hardware, and resources for data transmission and environmental sensing \cite{Liu2022ISAC}. In underwater acoustic systems, ISAC is particularly useful because bandwidth and energy are scarce. Recent underwater ISAC studies have demonstrated its potential for resource prioritization, environmental perception, and shallow-water channel monitoring \cite{Hazarika2025UWISAC}. However, integrating ISAC with TDOA localization introduces a topology management challenge: activating all links may improve localization accuracy but increases bandwidth usage, energy consumption, and delay, whereas using too few links can reduce Fisher information and degrade positioning accuracy. Thus, this letter reformulates topology-managed TDOA localization for ISAC-enabled IoUT networks by jointly considering acoustic ISAC link formation, sound-speed uncertainty, residual bias, multipath/Doppler-induced errors, link reliability, and node-level bandwidth, computation, and energy limitations.

Different from terrestrial IoT localization, the proposed ISAC-TM-MAE framework accounts for underwater acoustic propagation, heterogeneous IoUT nodes, and ISAC-enabled acoustic links \cite{Zheng2025UWAISAC}. The IoUT network is modeled as a directed graph whose edges represent ISAC-TDOA measurements and data-flow directions. For modeling rigor and fair comparison, all methods are evaluated under the same IoUT deployment, acoustic timing-noise, connectivity, and node-density settings, while the proposed topology selection explicitly accounts for residual bias, link reliability, and bandwidth--energy--computation constraints.

The main contributions of this letter are summarized as follows:
\begin{enumerate}
    \item We formulate an ISAC-enabled TDOA localization model for heterogeneous IoUT networks.
    \item We develop a resource-aware D-optimal topology selection strategy under acoustic bandwidth, energy, computation, and reliability constraints.
    \item We design an ISAC-TM-MAE estimator with compensated acoustic range differences and reliability-aware refinement.
    \item We compare the proposed method with centralized/distributed IoUT processing, UWA-CRLB~\cite{Gong2023UnderwaterLimits}, MDS-C~\cite{khalil2021bayesian}, MDS-D~\cite{Costa2006DistributedMDS}, and SDP~\cite{Wang2022SDP}, including robustness tests for node drift and acoustic link breakage.
\end{enumerate}
\vspace{-0.5em}

\section{System Model and Problem Formulation}\label{systemmodel}
\begin{figure}[t!]
    \centering
    \includegraphics[width=1\columnwidth]{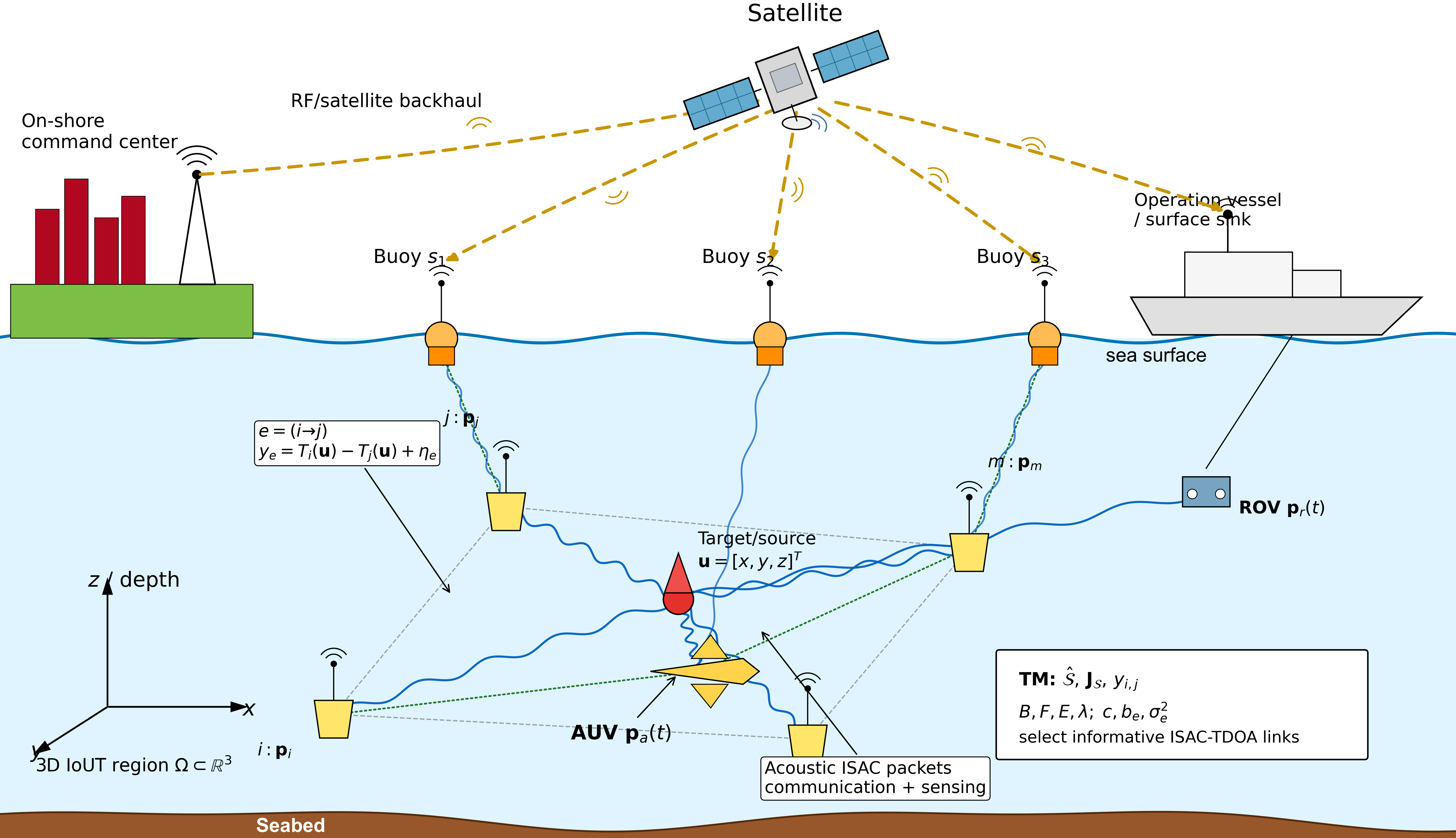}
    \caption{Proposed ISAC-enabled IoUT system model for topology-managed TDOA localization.}
    \label{fig:iout_isac_system_model}
\end{figure}

We consider an ISAC-enabled IoUT network deployed in a bounded three-dimensional underwater region $\Omega\subset\mathbb{R}^{3}$. 
The network contains $N$ cooperative nodes $\mathcal{V}=\{1,\ldots,N\}$, including seabed anchors, underwater sensors, AUVs, and surface buoys. 
The position of node $n$ is $\mathbf{p}_{n}=[x_n,y_n,z_n]^T$, while the unknown target/source location is $\mathbf{u}=[x,y,z]^T\in\Omega$. 
During each localization interval, acoustic ISAC packets are used for both communication and sensing-assisted delay estimation, reducing dedicated localization transmissions \cite{Liu2022ISAC,Zheng2025UWAISAC}. For a scheduled acoustic ISAC packet over link $e$, the transmitted waveform is modeled as
\begin{equation}
s_e(t)=\sqrt{P_e}\left(\sqrt{1-\alpha_e}\,d_e(t)+\sqrt{\alpha_e}\,p_e(t)\right),
\label{eq:isac_waveform}
\end{equation}
where $P_e$ is the transmit power, $d_e(t)$ carries data symbols, $p_e(t)$ is a known sensing pilot/chirp, and $\alpha_e$ controls the sensing--communication resource split. A receiver timing observation is extracted by correlating the received packet with $p_e(t)$, and the TDOA measurement is formed by differencing the timing observations of two receiving nodes. Thus, the ISAC waveform affects the timing variance $\sigma_e^2$, link bandwidth $\beta_e$, energy cost $\xi_{e,n}$, and reliability $\lambda_e$.
Here, simultaneous communication and localization refer to waveform/resource sharing within scheduled acoustic ISAC packets, not ideal full-duplex operation. Practical narrow-band and half-duplex transducer limitations are reflected through $\beta_e$, $B_n^{\mathrm{tx}}$, $B_n^{\mathrm{rx}}$, $\lambda_e$, and $\sigma_e^2$.
Fig.~\ref{fig:iout_isac_system_model} illustrates the considered topology-managed ISAC-enabled IoUT localization system.

Underwater acoustic propagation is affected by sound-speed variation, multipath, Doppler spread, and frequency-dependent attenuation \cite{Stojanovic2009UWAChannel}. 
Let $c(\mathbf{r};\boldsymbol{\chi})$ denote the sound speed at location $\mathbf{r}$, where $\boldsymbol{\chi}$ includes temperature, salinity, pressure, and depth. 
The acoustic travel time between $\mathbf{u}$ and node $n$ is
\begin{equation}
\small
    T_n(\mathbf{u};\boldsymbol{\chi})
    =
    \int_{\Gamma_n(\mathbf{u})}
    \frac{1}{c(\mathbf{r};\boldsymbol{\chi})}\,d\ell,
    \label{eq:travel_time}
\end{equation}
where $\Gamma_n(\mathbf{u})$ is the acoustic path. 
For moderate-depth deployments with limited ray bending,
\begin{equation}
\small
    T_n(\mathbf{u};\boldsymbol{\chi})
    \simeq
    \frac{\|\mathbf{u}-\mathbf{p}_n\|}{\bar{c}_n},
    \label{eq:travel_time_approx}
\end{equation}
where $\bar{c}_n$ is the path-averaged effective sound speed. 
The approximation in~\eqref{eq:travel_time_approx} is used only for tractable localization modeling and does not assume a perfectly homogeneous channel. Residual errors due to sound-speed mismatch and unmodeled ray bending are captured by the bias term $b_e$ and the SSP-related variance $\sigma_{\mathrm{ssp},e}^{2}$ in~\eqref{eq:uwa_noise}. If measured sound-speed profiles or ray-tracing models are available, the general travel time in~\eqref{eq:travel_time} can be directly used in the FIM and topology-selection procedure.

Let $\mathcal{E}$ be the set of candidate directed acoustic ISAC links. 
For link $e=(i\rightarrow j)\in\mathcal{E}$, the direction determines communication, energy, and processing load, while the TDOA feature is formed by nodes $i$ and $j$. 
The TDOA observation is modeled as
\begin{equation}
    y_e
    =
    T_i(\mathbf{u};\boldsymbol{\chi})
    -
    T_j(\mathbf{u};\boldsymbol{\chi})
    +
    b_e
    +
    \eta_e,
    \label{eq:iout_tdoa}
\end{equation}
where $b_e$ captures residual bias due to sound-speed-profile mismatch, unresolved multipath, and Doppler mismatch, and $\eta_e$ is a zero-mean timing error with variance
\begin{equation}
    \sigma_e^2
    =
    \sigma_{\mathrm{clk},e}^{2}
    +
    \sigma_{\mathrm{mp},e}^{2}
    +
    \sigma_{\mathrm{dop},e}^{2}
    +
    \sigma_{\mathrm{ssp},e}^{2}.
    \label{eq:uwa_noise}
\end{equation}
Here, the four terms denote clock/timing, multipath, Doppler-induced, and sound-speed-profile uncertainty components, respectively.

For a selected link set $\mathcal{S}\subseteq\mathcal{E}$, the stacked TDOA model is
\begin{equation}
    \mathbf{y}_{\mathcal{S}}
    =
    \mathbf{g}_{\mathcal{S}}(\mathbf{u})
    +
    \mathbf{b}_{\mathcal{S}}
    +
    \boldsymbol{\eta}_{\mathcal{S}},
    \label{eq:stacked_measurement}
\end{equation}
where $\mathbf{g}_{\mathcal{S}}(\mathbf{u})$ contains the noiseless travel-time differences, $\mathbf{b}_{\mathcal{S}}$ is the residual bias vector, and $\boldsymbol{\eta}_{\mathcal{S}}$ has covariance $\mathbf{R}_{\mathcal{S}}$. 
The corresponding FIM is
\begin{equation}
    \mathbf{J}_{\mathcal{S}}(\mathbf{u})
    =
    \mathbf{H}_{\mathcal{S}}^{T}(\mathbf{u})
    \mathbf{R}_{\mathcal{S}}^{-1}
    \mathbf{H}_{\mathcal{S}}(\mathbf{u}),
    \label{eq:fim_iout}
\end{equation}
where $\mathbf{H}_{\mathcal{S}}(\mathbf{u})$ is the Jacobian of $\mathbf{g}_{\mathcal{S}}(\mathbf{u})$ with respect to $\mathbf{u}$. 
For $e=(i\rightarrow j)$, the Jacobian row is
\begin{equation}
\small
    \mathbf{h}_{e}^{T}(\mathbf{u})
    =
    \nabla_{\mathbf{u}}
    \left[
    T_i(\mathbf{u};\boldsymbol{\chi})
    -
    T_j(\mathbf{u};\boldsymbol{\chi})
    \right].
    \label{eq:jacobian_general}
\end{equation}
Using~\eqref{eq:travel_time_approx}, this becomes
\begin{equation}
\small
    \mathbf{h}_{e}^{T}(\mathbf{u})
    \simeq
    \frac{(\mathbf{u}-\mathbf{p}_i)^T}
    {\bar{c}_i\|\mathbf{u}-\mathbf{p}_i\|}
    -
    \frac{(\mathbf{u}-\mathbf{p}_j)^T}
    {\bar{c}_j\|\mathbf{u}-\mathbf{p}_j\|}.
    \label{eq:jacobian_approx}
\end{equation}
Hence, the CRLB of any unbiased estimator of $\mathbf{u}$ is
\begin{equation}
    \mathrm{CRLB}_{\mathcal{S}}(\mathbf{u})
    =
    \mathbf{J}_{\mathcal{S}}^{-1}(\mathbf{u}).
    \label{eq:crlb_iout}
\end{equation}

Since $\mathbf{u}$ is unknown, $\Omega$ is discretized into grid points $\mathcal{U}=\{\mathbf{u}_1,\ldots,\mathbf{u}_L\}$. 
To avoid activating all TDOA links, a compact ISAC topology is selected by maximizing
\begin{equation}
\small
    \Phi(\mathcal{S})
    =
    \frac{1}{L}
    \sum_{\ell=1}^{L}
    \log\det
    \left(
    \mathbf{J}_{\mathcal{S}}(\mathbf{u}_{\ell})
    +
    \varepsilon\mathbf{I}_{3}
    \right),
    \label{eq:d_optimal_utility}
\end{equation}
where $\varepsilon>0$ prevents singularity for weak geometries.

Define $a_e=1$ if link $e$ is selected and $a_e=0$ otherwise. 
Let $\tau(e)$ and $\psi(e)$ denote the tail and head nodes of $e$, and let $\mathcal{S}(\mathbf{a})=\{e\in\mathcal{E}\mid a_e=1\}$. 
For link $e$, $\beta_e$ is the bandwidth cost, $\gamma_e$ is the processing load, $\xi_{e,n}$ is the energy consumed by node $n$, and $\lambda_e$ is the link reliability. 
The energy term $\xi_{e,n}$ is a generic link-dependent cost that may include transmit, receive, idle/listening, and processing energy. Thus, nonlinear modem-specific power profiles can be incorporated into $\xi_{e,n}$ using measured data or modem parameters; in the simulations, it is implemented as a normalized link cost.
Each node has transmit bandwidth $B_n^{\mathrm{tx}}$, receive bandwidth $B_n^{\mathrm{rx}}$, computational capacity $F_n$, and residual energy $E_n^{\mathrm{res}}$. 
The topology selection problem is
\begin{subequations}
\begin{align}
\small
    \max_{\{a_e\}_{e\in\mathcal{E}}}\quad
    &
    \Phi\big(\mathcal{S}(\mathbf{a})\big)
    \label{eq:optimization_main}
    \\
    \mathrm{s.t.}\quad
    &
    a_e\in\{0,1\},
    \qquad \forall e\in\mathcal{E},
    \label{eq:binary_constraint}
    \\
    &
    \sum_{e:\tau(e)=n} a_e\beta_e
    \leq
    B_n^{\mathrm{tx}},
    \qquad \forall n\in\mathcal{V},
    \label{eq:tx_constraint}
    \\
    &
    \sum_{e:\psi(e)=n} a_e\beta_e
    \leq
    B_n^{\mathrm{rx}},
    \qquad \forall n\in\mathcal{V},
    \label{eq:rx_constraint}
    \\
    &
    \sum_{e:\psi(e)=n} a_e\gamma_e
    \leq
    F_n,
    \qquad \forall n\in\mathcal{V},
    \label{eq:compute_constraint}
    \\
    &
    \sum_{e\in\mathcal{E}_n} a_e\xi_{e,n}
    \leq
    E_n^{\mathrm{res}},
    \qquad \forall n\in\mathcal{V},
    \label{eq:energy_constraint}
    \\
    &
    a_e\lambda_e
    \geq
    a_e\lambda_{\min},
    \qquad \forall e\in\mathcal{E},
    \label{eq:reliability_constraint}
    \\
    &
    K_{\min}
    \leq
    \sum_{e\in\mathcal{E}} a_e
    \leq
    K_{\max}.
    \label{eq:cardinality_constraint}
\end{align}
\end{subequations}
Here, $\mathcal{E}_n$ is the set of links incident to node $n$, $\lambda_{\min}$ is the minimum acceptable reliability, and $K_{\min}$ and $K_{\max}$ bound the active topology size. 
The constraints limit acoustic signaling, computation, energy, and unreliable links, while the objective favors TDOA geometries with improved localization.
\vspace{-1.0em}

\section{Proposed ISAC-TM-MAE Method}\label{maealg}
The proposed method selects a reliable ISAC-TDOA topology and refines the estimate via local fusion and residual correction, reducing transmissions under acoustic uncertainty and energy-limited signaling \cite{Stojanovic2009UWAChannel}. 
Unlike direct terrestrial D-optimal FIM/WLS use, it handles residual bias, sound-speed mismatch, multipath/Doppler errors, and link reliability through compensated range differences, resource-aware selection, and reliability-aware correction. Its three-stage design avoids nonlinear WLS by providing stable initialization, enforcing hub-range consistency, and applying local refinement, lowering complexity for resource-constrained IoUT nodes.

Let $\mathcal{A}$ be the feasible set of binary link-selection vectors satisfying the resource and reliability constraints, and let $\mathcal{U}=\{\mathbf{u}_{\ell}\}_{\ell=1}^{L}$ be the localization grid. The selected topology is obtained using the D-optimal criterion
\begin{equation}
\small
    \hat{\mathbf{a}}
    =
    \arg\max_{\mathbf{a}\in\mathcal{A}}
    \frac{1}{L}
    \sum_{\ell=1}^{L}
    \log\det
    \left(
    \mathbf{J}_{\mathbf{a}}(\mathbf{u}_{\ell})
    +
    \varepsilon\mathbf{I}_{3}
    \right),
    \label{eq:mae_topology_selection}
\end{equation}
where $\mathbf{J}_{\mathbf{a}}(\mathbf{u}_{\ell})$ is the FIM for topology $\mathbf{a}$ and $\varepsilon>0$ is a regularization factor. 
The selected link set is
\begin{equation}
    \hat{\mathcal{S}}
    =
    \{e\in\mathcal{E}\mid \hat{a}_{e}=1\}.
\end{equation}
The graph $\hat{\mathcal{S}}$ is then divided into $M$ local fusion groups, where $\varrho_m$ denotes the hub of the $m$-th group and $\mathcal{N}_{m}$ denotes its neighboring nodes.

For each selected pair $(\varrho_m,n)$, the ISAC-TDOA observation is converted into a compensated acoustic range difference as
\begin{equation}
    \delta_{m,n}
    =
    \varsigma_{m,n}\hat{c}_{m,n}
    \left(
    y_{m,n}
    -
    \hat{b}_{m,n}
    \right),
    \label{eq:range_difference}
\end{equation}
where $\hat{c}_{m,n}$ is the estimated effective sound speed, $\hat{b}_{m,n}$ is the residual bias compensation, and $\varsigma_{m,n}\in\{-1,1\}$ accounts for link direction. 
Thus,
\begin{equation}
    \delta_{m,n}
    =
    \|\mathbf{u}-\mathbf{p}_{n}\|
    -
    \|\mathbf{u}-\mathbf{p}_{\varrho_m}\|
    +
    \nu_{m,n},
    \label{eq:compensated_tdoa}
\end{equation}
where $\nu_{m,n}$ represents the residual acoustic error.

In the first stage, introduce the hub-range variable
$r_m=\|\mathbf{u}-\mathbf{p}_{\varrho_m}\|$ and define
\begin{equation}
    \boldsymbol{\vartheta}_{1}
    =
    \left[
    \mathbf{u}^{T},
    r_{1},
    \ldots,
    r_{M}
    \right]^{T}.
\end{equation}
From \eqref{eq:compensated_tdoa}, the following pseudo-linear relation is obtained:
\begin{equation}
    \mathbf{a}_{m,n}^{T}\boldsymbol{\vartheta}_{1}
    =
    \ell_{m,n}
    +
    \zeta_{m,n},
    \label{eq:first_stage_linear}
\end{equation}
with
\begin{align}
    \mathbf{a}_{m,n}^{T}
    &=
    2
    \left[
    (\mathbf{p}_{n}-\mathbf{p}_{\varrho_m})^{T},
    \mathbf{0}_{1\times(m-1)},
    \delta_{m,n},
    \mathbf{0}_{1\times(M-m)}
    \right],
    \label{eq:first_stage_row}
    \\
    \ell_{m,n}
    &=
    \|\mathbf{p}_{n}\|^{2}
    -
    \|\mathbf{p}_{\varrho_m}\|^{2}
    -
    \delta_{m,n}^{2}.
    \label{eq:first_stage_rhs}
\end{align}
Stacking all selected measurements gives
\begin{equation}
    \mathbf{A}_{1}\boldsymbol{\vartheta}_{1}
    =
    \boldsymbol{\ell}_{1}
    +
    \boldsymbol{\zeta}_{1}.
\end{equation}
An initial regularized least-squares (LS) estimate is computed as
\begin{equation}
    \hat{\boldsymbol{\vartheta}}_{1}^{(0)}
    =
    \left(
    \mathbf{A}_{1}^{T}\mathbf{A}_{1}
    +
    \mu\mathbf{I}
    \right)^{-1}
    \mathbf{A}_{1}^{T}
    \boldsymbol{\ell}_{1},
    \label{eq:first_stage_ls}
\end{equation}
and the refined first-stage weighted LS (WLS) estimate is
\begin{equation}
    \hat{\boldsymbol{\vartheta}}_{1}
    =
    \left(
    \mathbf{A}_{1}^{T}\mathbf{W}_{1}\mathbf{A}_{1}
    +
    \mu\mathbf{I}
    \right)^{-1}
    \mathbf{A}_{1}^{T}\mathbf{W}_{1}
    \boldsymbol{\ell}_{1},
    \label{eq:first_stage_wls}
\end{equation}
where $\mathbf{W}_{1}=(\mathbf{D}_{1}\mathbf{Q}_{\delta}\mathbf{D}_{1}^{T}+\varepsilon_w\mathbf{I})^{-1}$. 
Let $K_s$ be the number of selected compensated range-difference measurements and define 
$\hat{\mathbf d}=[\hat d_{m_1,n_1},\ldots,\hat d_{m_{K_s},n_{K_s}}]^T$, where 
$\hat d_{m_k,n_k}=\|\hat{\mathbf u}^{(0)}-\mathbf p_{n_k}\|$. 
Then, $\mathbf{D}_{1}=\mathrm{diag}(2\hat{\mathbf d})\in\mathbb{R}^{K_s\times K_s}$.

\begin{figure*}[t]
    \centering
    \subfloat[15 topology updates.]
    {
        \includegraphics[width=0.31\textwidth]{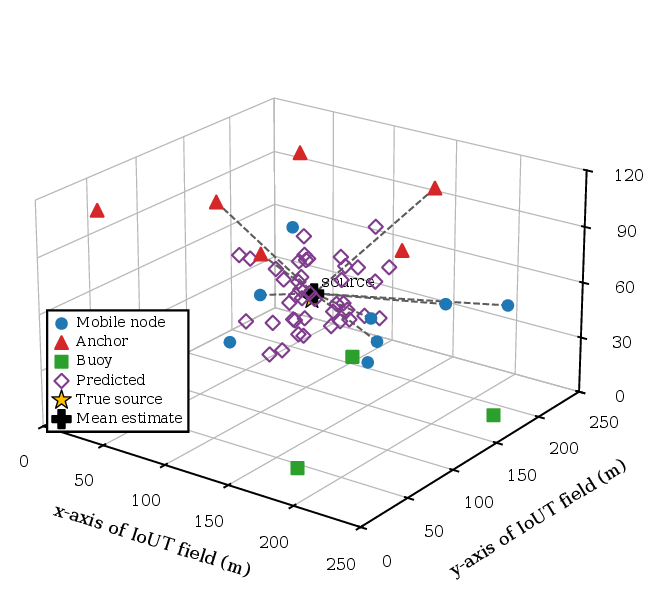}
        \label{fig:iout_3d_stage1}
    }
    \hfill
    \subfloat[30 topology updates.]
    {
        \includegraphics[width=0.31\textwidth]{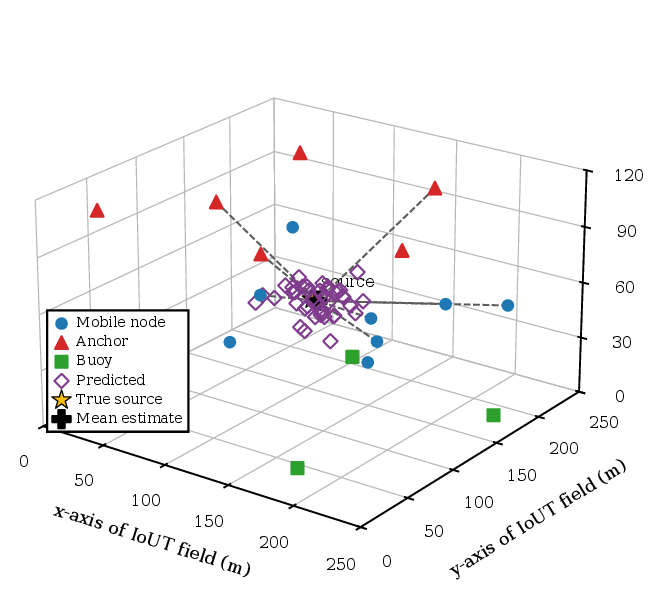}
        \label{fig:iout_3d_stage2}
    }
    \hfill
    \subfloat[45 topology updates.]
    {
        \includegraphics[width=0.31\textwidth]{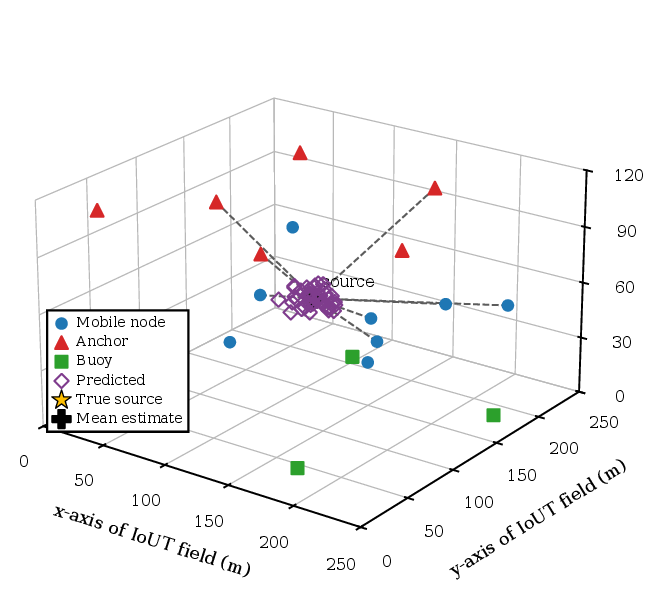}
        \label{fig:iout_3d_stage3}
    }
    \caption{IoUT localization using the proposed ISAC-TM-MAE framework after 15, 30, and 45 topology updates.}
    \label{fig:iout_3d_localization_subfigures}
\end{figure*}

The second stage enforces consistency between the target position and the estimated hub ranges. 
Let $\hat{r}_{m}$ be extracted from $\hat{\boldsymbol{\vartheta}}_{1}$, introduce $\kappa=\mathbf{u}^{T}\mathbf{u}$, and define
$\boldsymbol{\vartheta}_{2}=[\mathbf{u}^{T},\kappa]^{T}$. 
For each hub,
\begin{equation}
    \mathbf{a}_{2,m}^{T}\boldsymbol{\vartheta}_{2}
    =
    \ell_{2,m}
    +
    \zeta_{2,m},
    \qquad
    \mathbf{a}_{2,m}^{T}
    =
    [-2\mathbf{p}_{\varrho_m}^{T},1],
\end{equation}
where $\ell_{2,m}=\hat{r}_{m}^{2}-\|\mathbf{p}_{\varrho_m}\|^{2}$. 
Stacking all hubs yields
$\mathbf{A}_{2}\boldsymbol{\vartheta}_{2}=\boldsymbol{\ell}_{2}+\boldsymbol{\zeta}_{2}$, and
\begin{equation}
    \hat{\boldsymbol{\vartheta}}_{2}
    =
    \left(
    \mathbf{A}_{2}^{T}\mathbf{W}_{2}\mathbf{A}_{2}
    +
    \mu\mathbf{I}
    \right)^{-1}
    \mathbf{A}_{2}^{T}\mathbf{W}_{2}
    \boldsymbol{\ell}_{2},
    \qquad
    \tilde{\mathbf{u}}
    =
    [\hat{\boldsymbol{\vartheta}}_{2}]_{1:3},
    \label{eq:second_stage_wls}
\end{equation}
where $\mathbf{W}_{2}$ is formed from the covariance of the first-stage hub-range estimates.

Finally, a reliability-aware nonlinear correction is applied using the original ISAC-TDOA model. 
The residual at $\tilde{\mathbf{u}}$ is
\begin{equation}
    \mathbf{r}_{3}
    =
    \mathbf{y}_{\hat{\mathcal{S}}}
    -
    \hat{\mathbf{b}}_{\hat{\mathcal{S}}}
    -
    \mathbf{g}_{\hat{\mathcal{S}}}
    \left(
    \tilde{\mathbf{u}};
    \hat{\boldsymbol{\chi}}
    \right),
    \label{eq:third_stage_residual}
\end{equation}
where $\mathbf{g}_{\hat{\mathcal{S}}}(\cdot)$ is the predicted TDOA vector. 
Using the Jacobian
$\mathbf{H}_{3}
=
\partial\mathbf{g}_{\hat{\mathcal{S}}}(\mathbf{u};\hat{\boldsymbol{\chi}})/\partial\mathbf{u}
|_{\mathbf{u}=\tilde{\mathbf{u}}}$,
the correction is
\begin{equation}
    \Delta\hat{\mathbf{u}}
    =
    \left(
    \mathbf{H}_{3}^{T}\mathbf{W}_{3}\mathbf{H}_{3}
    +
    \mu\mathbf{I}_{3}
    \right)^{-1}
    \mathbf{H}_{3}^{T}\mathbf{W}_{3}
    \mathbf{r}_{3},
    \label{eq:third_stage_correction}
\end{equation}
where 
$\mathbf{W}_{3}
=
\boldsymbol{\Lambda}
(\mathbf{R}_{\hat{\mathcal{S}}}+\mathbf{R}_{b}+\mathbf{R}_{\chi})^{-1}
\boldsymbol{\Lambda}$
weights the selected links according to measurement noise, bias uncertainty, environmental uncertainty, and link reliability. 
The final estimate is
\begin{equation}
    \hat{\mathbf{u}}
    =
    \Pi_{\Omega}
    \left(
    \tilde{\mathbf{u}}
    +
    \Delta\hat{\mathbf{u}}
    \right),
    \label{eq:final_iout_estimate}
\end{equation}
where $\Pi_{\Omega}(\cdot)$ projects the solution onto the feasible underwater deployment region.

The complete procedure is summarized in Algorithm~\ref{alg:iout_mae}.
\begin{algorithm}[h!]
\scriptsize
\caption{Proposed ISAC-TM-MAE for IoUT Localization.}
\label{alg:iout_mae}
\begin{algorithmic}[1]
\Require $\{\mathbf{p}_{n}\}_{n=1}^{N}$, $\mathcal{E}$, $\{y_e\}$, $\hat{\boldsymbol{\chi}}$, constraints, $\mathcal{U}$
\Ensure $\hat{\mathcal{S}}$, $\hat{\mathbf{u}}$

\State Select $\hat{\mathbf{a}}$ by evaluating $\mathbf{J}_{\mathbf{a}}(\mathbf{u}_{\ell})$ over feasible $\mathbf{a}$ and $\mathbf{u}_{\ell}\in\mathcal{U}$ using~\eqref{eq:mae_topology_selection}.
\State Form $\hat{\mathcal{S}}=\{e\in\mathcal{E}\mid \hat{a}_{e}=1\}$ and divide it into hub-based local fusion groups.
\State Convert selected ISAC--TDOA observations to compensated range differences $\delta_{m,n}$ using~\eqref{eq:range_difference}.
\State Estimate $\hat{\boldsymbol{\vartheta}}_{1}$ from the first-stage WLS system $\mathbf{A}_{1}\boldsymbol{\vartheta}_{1}=\boldsymbol{\ell}_{1}$.
\State Obtain $\tilde{\mathbf{u}}$ from the hub-range consistency WLS system $\mathbf{A}_{2}\boldsymbol{\vartheta}_{2}=\boldsymbol{\ell}_{2}$.
\State Refine $\tilde{\mathbf{u}}$ using the nonlinear residual correction $\Delta\hat{\mathbf{u}}$.
\State Output $\hat{\mathbf{u}}=\Pi_{\Omega}(\tilde{\mathbf{u}}+\Delta\hat{\mathbf{u}})$ and $\hat{\mathcal{S}}$.
\end{algorithmic}
\end{algorithm}
\vspace{-1em}

\section{Results and Analysis}\label{results}
The simulations are conducted in a 3D IoUT region $\Omega=[0,250]\times[0,250]\times[0,120]~\mathrm{m}^{3}$ with seabed anchors, surface buoys, and mobile AUV/ROV receivers localizing an underwater acoustic source. The effective sound speed is $\bar{c}=1500~\mathrm{m/s}$, while the timing jitter, connectivity radius, and number of cooperative receivers are varied as $\sigma_{\tau}\in[0.10,1.30]~\mathrm{ms}$, $R_a\in[120,720]~\mathrm{m}$, and $N\in[12,60]$, respectively. The selected topology is obtained using the proposed D-optimal FIM-based criterion under $\{B_n^{\mathrm{tx}},B_n^{\mathrm{rx}},F_n,E_n^{\mathrm{res}},\lambda_e\}$, and the accuracy is measured by $\mathrm{RMSE}=\sqrt{\frac{1}{N_{\mathrm{mc}}}\sum_{\varpi=1}^{N_{\mathrm{mc}}}\|\hat{\mathbf{u}}^{(\varpi)}-\mathbf{u}^{(\varpi)}\|^{2}}$. The proposed scheme is compared with centralized/distributed IoUT processing~\cite{Gong2023UnderwaterLimits}, UWA-CRLB, MDS-C~\cite{khalil2021bayesian}, MDS-D~\cite{Costa2006DistributedMDS}, and SDP~\cite{Wang2022SDP}. 
For dynamic IoUT evaluation, passive node drift is modeled as $\tilde{\mathbf{p}}_n=\mathbf{p}_n+\Delta\mathbf{p}_n$, where $\Delta\mathbf{p}_n\sim\mathcal{N}(\mathbf{0},\sigma_d^2\mathbf{I}_3)$, and link breakage is modeled by $\omega_e\in\{0,1\}$ with probability $p_b$, so topology selection is performed over $\mathcal{E}_a=\{e\in\mathcal{E}\mid \omega_e=1\}$. The timing disturbance follows zero-mean Gaussian noise with variance $\sigma_e^2$ in~\eqref{eq:uwa_noise}, while $b_e$ captures deterministic residual bias; experimental validation with measured underwater acoustic ISAC data is left for future work.

Fig.~\ref{fig:iout_3d_localization_subfigures} shows the 3D localization behavior of the proposed ISAC-TM-MAE after 15, 30, and 45 topology updates. The predicted source locations are initially dispersed due to underwater timing errors, multipath, Doppler distortion, and sound-speed uncertainty, but become increasingly concentrated around the true acoustic source as the topology and MAE refinement progress. This confirms that the proposed framework improves localization stability while avoiding activation of all IoUT links.

\begin{figure*}[h!]
    \centering
    \subfloat[Noise standard deviation.]
    {
        \includegraphics[width=0.31\textwidth]{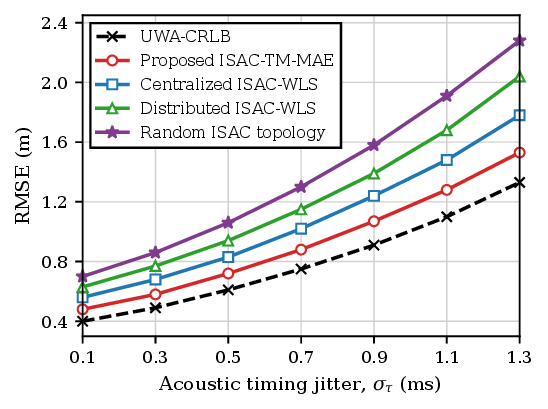}
        \label{fig:rmse_noise}
    }
    \hfill
    \subfloat[Connectivity range.]
    {
        \includegraphics[width=0.31\textwidth]{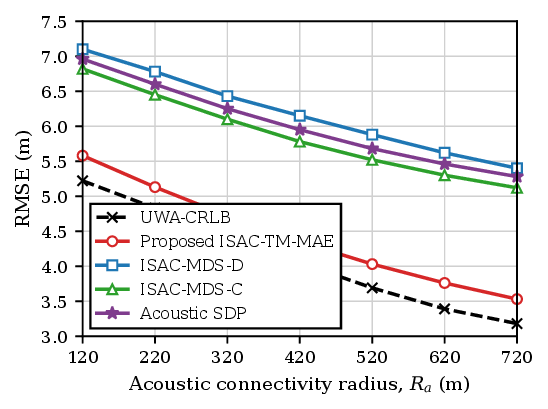}
        \label{fig:rmse_connectivity}
    }
    \hfill
    \subfloat[Number of IoUT nodes.]
    {
        \includegraphics[width=0.31\textwidth]{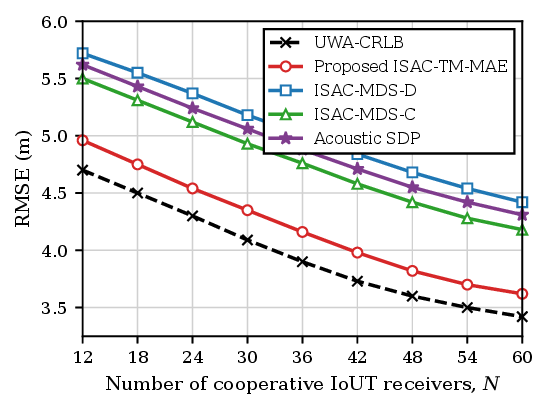}
        \label{fig:rmse_nodes}
    }
    \\[-1mm]
    \subfloat[Node drift.]
    {
        \includegraphics[width=0.31\textwidth]{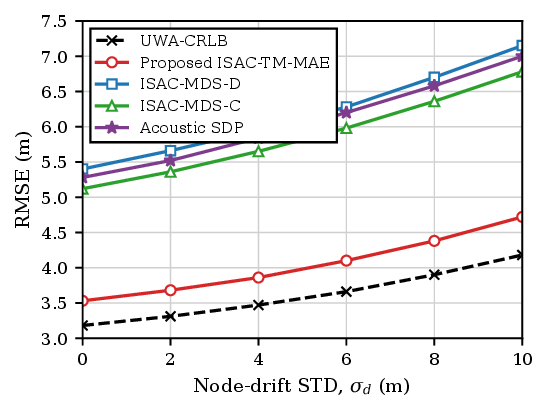}
        \label{fig:node_drift}
    }
    \hfill
    \subfloat[Link breakage.]
    {
        \includegraphics[width=0.31\textwidth]{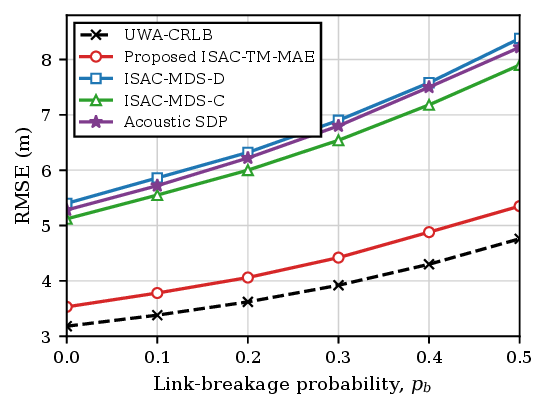}
        \label{fig:link_breakage}
    }
    \hfill
    \subfloat[Topology ablation.]
    {
        \includegraphics[width=0.31\textwidth]{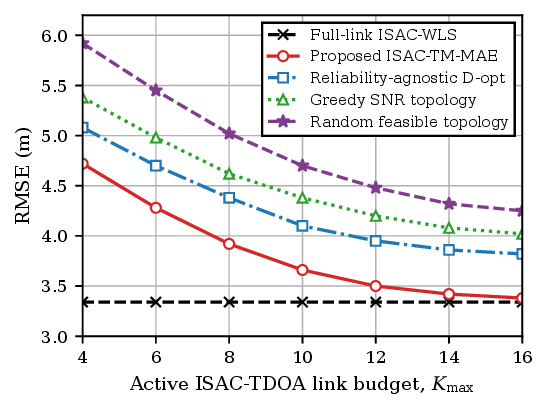}
        \label{fig:topology_ablation}
    }
    \caption{RMSE performance of the proposed ISAC-TM-MAE framework under different IoUT localization, robustness, and topology-selection scenarios.}
    \label{fig:all_rmse_results}
\end{figure*}

Figs.~\ref{fig:rmse_noise}--\ref{fig:rmse_nodes} show the RMSE performance under timing noise, connectivity range, and node-density variations, compared with centralized/distributed IoUT processing~\cite{Gong2023UnderwaterLimits}, UWA-CRLB~\cite{Gong2023UnderwaterLimits}, MDS-C~\cite{khalil2021bayesian}, MDS-D~\cite{Costa2006DistributedMDS}, and SDP~\cite{Wang2022SDP}. As $\sigma_{\tau}$ increases from $0.10$ ms to $1.30$ ms, the proposed ISAC-TM-MAE increases from about $0.48$ m to $1.53$ m, while centralized, distributed, and random topology schemes reach about $1.78$ m, $2.04$ m, and $2.28$ m, respectively. Increasing $R_a$ from $120$ m to $720$ m reduces the proposed RMSE from $5.58$ m to $3.53$ m, which is about $35\%$ lower than ISAC-MDS-D~\cite{Costa2006DistributedMDS} at $R_a=720$ m. Similarly, increasing $N$ from $12$ to $60$ reduces the proposed RMSE from $4.96$ m to $3.62$ m, remaining close to the UWA-CRLB of $3.42$ m~\cite{Gong2023UnderwaterLimits}. These results suggest that the proposed topology model can effectively exploit informative acoustic links with reduce unnecessary measurements.

Figs.~\ref{fig:node_drift} and~\ref{fig:link_breakage} evaluate robustness under dynamic IoUT conditions. When $\sigma_d$ increases from $0$ m to $10$ m, the proposed method increases from about $3.53$ m to $4.72$ m, while ISAC-MDS-D~\cite{Costa2006DistributedMDS} and acoustic SDP~\cite{Wang2022SDP} reach about $7.15$ m and $7.00$ m, respectively. When $p_b=0.5$, the proposed method achieves about $5.35$ m RMSE, compared with $8.38$ m, $7.90$ m, and $8.22$ m for ISAC-MDS-D~\cite{Costa2006DistributedMDS}, ISAC-MDS-C~\cite{khalil2021bayesian}, and acoustic SDP~\cite{Wang2022SDP}, respectively. Fig.~\ref{fig:topology_ablation} further shows that at $K_{\max}=8$, the proposed ISAC-TM-MAE achieves about $3.92$ m RMSE, compared with $4.38$ m, $4.62$ m, and $5.02$ m for reliability-agnostic D-optimal, greedy SNR, and random topology selection, respectively. These ablation baselines isolate the effect of resource-aware link selection and show that the proposed method provides a better accuracy--resource tradeoff by selecting reliable acoustic links without activating all available IoUT links.

\section{Conclusion}\label{conclusion}
This letter presents the proposed ISAC-TM-MAE framework for TDOA localization in IoUT networks. By integrating D-optimal topology management with multi-stage adaptive estimation, the proposed framework selects informative and reliable acoustic links while accounting for bandwidth, energy, computation, and link-reliability constraints. Numerical results show that the proposed method improves localization accuracy compared with centralized, distributed, MDS-C, MDS-D, and SDP-based methods, while closely approaching the underwater acoustic CRLB. The additional robustness results further demonstrate its effectiveness under node drift, acoustic link breakage, and limited active-link budgets. Therefore, the proposed method provides a resource-efficient and robust localization solution for underwater ISAC-enabled IoUT networks.
\vspace{-1em}

\bibliographystyle{IEEEtran}
\bibliography{mybibliography}

\end{document}